\documentclass[twoside,english,reqno]{amsart}
\usepackage[T1]{fontenc}
\usepackage{amstext}
\usepackage{amsthm}

\usepackage{fancyhdr}
\makeatletter
\numberwithin{equation}{section}
\usepackage[colorlinks=true, linkcolor=red, citecolor=blue]{hyperref}

\usepackage[numbers,sort&compress]{natbib}
    {\begin{subequations}
        }%
    {\end{subequations}}

\makeatother

\usepackage{babel}

\newcommand{\ls}{\setlength{\baselineskip}{12pt}
                 \setlength{\parskip}{3mm}}

\begin{document}
\allowdisplaybreaks

\title{Nonexistence of Majorana fermions in Kerr-Newman type spacetimes with nontrivial charge}
\author[H Zhang]{He-Qun Zhang $^{2}$}
\author[X Zhang]{Xiao Zhang $^{1,2}$}
\address[]{$^{1}$Academy of Mathematics and Systems Science, Chinese Academy of Sciences, Beijing 100190, PR China}
\address[]{$^{2}$School of Mathematical Sciences, University of Chinese Academy of Sciences, Beijing 100049, PR China}
\email{zhanghequn@amss.ac.cn$^{2}$}
\email{xzhang@amss.ac.cn$^{1,2}$}
\begin{abstract}
We show that the Dirac equation is separated into four differential equations for time-period Majarana fermions in Kerr-Newman, Kerr-Newman-(A)dS spacetimes. Although they can not be transformed into radial and angular equations, the four differential equations yield two algebraic identities. When the electric or magnetic charge is nonzero, they conclude that there is no differentiable time-periodic Majorana fermions outside the event horizon in Kerr-Newman and Kerr-Newman-AdS spacetimes, or between the event horizon and the cosmological horizon in Kerr-Newman-dS spacetime.
\end{abstract}

\maketitle

\section{Introduction}
\ls

The Dirac equation in black hole spacetimes plays a significant role in the study of general relativity and quantum cosmology. A Dirac fermion is a spinor $\Psi$ in spacetimes satisfying the Dirac equation
\begin{equation}
    \big(D+i\lambda \big)\Psi=0,
    \label{eq:Diracequ}
\end{equation}
where $\lambda$ is certain real number. In 1968, Kinnersley introduced null basis to study the Petrov type D metric \cite{Kinnersley1969}. In \cite{Chandrasekhar1976}, Chandrasekhar separated the Dirac equation in Kerr spacetime when Dirac fermions are time-period and given by
\begin{align}\label{Dirac separation}
\Psi=S^{-1}\psi,\quad
\psi=e^{-i(\omega t+(k+\frac{1}{2})\phi)} \left(\begin{array}{c}
R_{-}(r)\Theta_{-}(\theta)\\
R_{+}(r)\Theta_{+}(\theta)\\
R_{+}(r)\Theta_{-}(\theta)\\
R_{-}(r)\Theta_{+}(\theta)
\end{array}\right),
\end{align}
where $S$ is a diagonal matrix
\begin{align*} S&=\Delta_{r}^{\ \frac{1}{4}}\text{diag}\left((r+ia\cos\theta)^{\frac{1}{2}}I_{2\times2},\,\,\,\,(r-ia\cos\theta)^{\frac{1}{2}}I_{2\times2}\right),
\end{align*}
and Page generated his method to Kerr-Newman spacetime \cite{page1976}. Since then, various works have been done to discuss the Hawking radiation and the numerical solutions of Dirac fermions in various spacetime backgrounds, see e.g. \cite{zhaozheng1981,Belgino1999,chakrabarti20002,Angheben2005}. In \cite{Felix2000}, Finster, Kamran, Smoller and Yau applied Chandrasekhar's separation to prove nonexistence of $L^2$ integrable, time-periodic solutions for the Dirac equation in non-extreme Kerr-Newman spacetime. This indicates that the normalizable time-periodic Dirac fermions must either disappear into the black hole or escape to infinity . In \cite{BC2009, Belgiorno2010}, Belgiorno and Cacciatori applied the spectral properties to prove non-existence of $L^2$ integrable, time-periodic solutions for the Dirac equation with mass greater than $\frac{1}{2} \sqrt{\frac{|\Lambda|}{3}}$ in non-extreme Kerr-Newman-(A)dS spacetimes, where $\Lambda$ is the cosmological constant. In \cite{zx2018}, Wang and Zhang applied Chandrasekhar's separation to prove the nonexistence of $L^p$ integrable, time-periodic solutions for the Dirac equation with arbitrary mass and $0<p\leq \frac{4}{3}$, or with mass greater than $q \sqrt{-\frac{\Lambda}{3}}$ and $\frac{4}{3}<p\leq \frac{4}{3-2q}$, $0<q<\frac{3}{2}$ in non-extreme Kerr-Newman-AdS spacetime. In non-extreme Kerr-Newman-dS spacetime, the nonexistence of $L^p$ integrable, time-periodic Dirac fermions hold true for arbitrary mass and $p\geq 2$ \cite{FWZ2024}. In particular, taking $p=2$, they recovered Belgiorno and Cacciatori's results that normalizable time-periodic Dirac fermions with mass greater than $\frac{1}{2} \sqrt{-\frac{\Lambda}{3}}$ must either disappear into the black hole or escape to infinity. 

For Chandrasekhar's separation, the Dirac equation can be transformed into radial and angular equations. In \cite{Kraniotis2019}, Kraniotis observed the radial and angular equations can be reduced to generalized Heun's equations in Kerr-Newman spacetime, which provide local time-periodic solutions in terms of holomorphic functions whose power series coefficients are determined by a four-term recurrence relation. Using the four-term recursion formula, he also proved that there is no time-period solution with fermion's energy strictly less than its mass in Kerr-Newman spacetime.

For the recent search for neutrinos, which is one of the most mysterious particles in the universe, we are interested in knowing whether they are Majorana fermions or not. The most promising way so far is through the double beta decay \cite{Rodejohann2011}. Various approaches were studied to distinguish the Majorana and Dirac fermions, see e.g. \cite{Alavi2012-DM, Cheng1980-DM, Garavaglia1983-DM, Nieves1987-DM, Rosen1982-DM, Shrock1982-DM, Singh2006-DM, Kim2021-DM}.

A Majorana fermion is a Dirac fermion whose antiparticle is itself. To define a Majarana fermion precisely, let us introduce the 4-component charge conjugate operator
\begin{equation*}
	C=\left(\begin{array}{cc}
\epsilon_{\beta\alpha}\\
& \epsilon^{\dot{\beta}\dot{\alpha}}
\end{array}\right)
\end{equation*}
with the Pauli matrix $\sigma_2$ and the antisymmetric operator on spin indices, where
\begin{equation*}
	\epsilon^{\dot{\alpha}\dot{\beta}}=-\epsilon_{\alpha\beta}=i\sigma_2=\left(\begin{array}{cc}
 & 1\\
-1
\end{array}\right).
\end{equation*}
The charge conjugation of the Dirac fermion $\Psi$ is defined by
\begin{equation*}
	\Psi^C=C\bar\Psi^T.
\end{equation*}
Therefore, Majorana fermions are given by
\begin{equation}\label{eq:majform}
	\Psi_{\text{Maj}}=\left(\begin{array}{c}
\Psi_{\text{Weyl}} \\
i\sigma_{2}\Psi_{\text{Weyl}}   ^{*}
\end{array}\right)
\end{equation}
and satisfies the Dirac equation \cite{Majorana1937, PU, HMH}, where $\Psi_{\text{Weyl}}$ is the Weyl spinor, and $\Psi_{\text{Weyl}} ^*$ is its complex conjugate. Time-period Majorana fermions can be given by
\begin{align}\label{majorana}
\Psi=S^{-1}E\psi,\quad
\psi=\left(\begin{array}{c}
R_{-}(r)\Theta_{-}(\theta)\\
R_{+}(r)\Theta_{+}(\theta)\\
\bar{R}_{+}(r)\bar{\Theta}_{+}(\theta)\\
-\bar{R}_{-}(r)\bar{\Theta}_{-}(\theta)
\end{array}\right),
\end{align}
where $S$ $,E$ are diagonal matrices
\begin{align*} 
S&=\Delta_{r}^{\ \frac{1}{4}}\text{diag}\left((r+ia\cos\theta)^{\frac{1}{2}}I_{2\times2},\,\,\,\,(r-ia\cos\theta)^{\frac{1}{2}}I_{2\times2}\right),\\
E&=\text{diag}\left(e^{-i(\omega t+(k+\frac{1}{2})\phi)}I_{2\times2},\,\,\,\,e^{i(\omega t+(k+\frac{1}{2})\phi)}I_{2\times2}\right).
\end{align*}

In this short paper, we show that the Dirac equation is separated into four differential equations for time-period Majarana fermions given by \eqref{majorana} in Kerr-Newman, Kerr-Newman-(A)dS spacetimes. Although they can not be transformed into radial and angular equations, the four different equations yield two algebraic identities. When the electric or magnetic charge is nonzero, they conclude that there is no differentiable time-periodic Majorana fermions outside the event horizon in Kerr-Newman and Kerr-Newman-AdS spacetimes, or between the event horizon and the cosmological horizon in Kerr-Newman-dS spacetime. 

We remark that Dirac fermions taking form \eqref{Dirac separation} are not consistent with the Majorana condition \eqref{eq:majform}. Thus previous results on the existence or non-existence of time-period Dirac fermions can not applied to the current situation for time-period Majorana fermions.

\section{Geometry of Kerr-Newman-type spacetimes}
\ls

For convenience of discussion, we unify Kerr-Newman, Kerr-Newman-(A)dS metrics
\begin{equation}\label{kerr}
\begin{aligned}
    ds_{\text{KNType}}^{2}=&-\frac{\Delta_{r}}{U}\left(dt-\frac{a\sin^{2}\theta}{\Xi}d\phi\right)^{2}
    +\frac{U}{\Delta_{r}}dr^{2}+\frac{U}{\Delta_{\theta}}d\theta^{2}\\
    &+\frac{\Delta_{\theta}\sin^{2}\theta}{U}\left(adt-\frac{r^{2}+a^{2}}{\Xi}d\phi\right)^{2},
\end{aligned}
\end{equation}
by taking $\kappa$ as zero, pure imaginary and real respectively, where $\Lambda=-3\kappa^{2}$ is the cosmological constant, and
\begin{align*}
    \Delta_{r}&=(r^{2}+a^{2})(1+\kappa^{2} r^{2}) -2mr+P^{2}+Q^{2},\\
    \Delta_{\theta}&=1-\kappa^{2} a^{2}\cos^{2}\theta,\quad 
    U=r^{2}+a^{2}\cos^{2}\theta,\quad
    \Xi = 1-\kappa^{2}a^{2}>0.
\end{align*}
The metric \eqref{kerr} solves the Einstein-Maxwell field equations with the electromagnetic potential
\begin{equation}
A=-\frac{Qr}{{U}}\left(dt-\frac{a\sin^2{\theta}}{\Xi}d{\varphi}\right)
-\frac{P\cos{\theta}}{U}\left(a dt-\frac{r^2+a^2}{\Xi}d\varphi\right) \label{A},
\end{equation}
where $P$, $Q$ are real numbers representing the magnetic charge and the electric charge respectively.

In the following we let $0 \leq \mu, \nu \leq 3$, and $1 \leq i, j\leq 3$. On a 4-dimensional Lorentzian manifold, we choose frame $\{e_\mu \}$ such that $e_0$ is timelike and $e_i$ are spacelike. Denote $\{e^\alpha \}$ the dual coframe. The Cartan structure equations are
\begin{equation*}
	de^{\mu}=-\omega^{\mu}_{\ \nu}\wedge e^\nu, \quad \omega_{\mu \nu}=g_{\mu \gamma} \omega^{\gamma}_{\ \nu}=-\omega_{\nu \mu}.
\end{equation*}
If it is spin, we use the cotangent bundle to define the Clifford multiplication, the spin connection and the Dirac operator \cite{zx2018, FWZ2024}. We fix the Clifford multiplications to be the following Weyl representation
\begin{equation}\label{clifford}
 e^{0}\mapsto\left(\begin{array}{cc}
 & I_{2\times2}\\
I_{2\times2}
\end{array}\right),\quad  e^{i}\mapsto\left(\begin{array}{cc}
 & \sigma_{i}\\
-\sigma_{i}
\end{array}\right),
\end{equation}
where $\sigma_i$ are Pauli matrices
\begin{align*}
	\sigma_{1}=\left(\begin{array}{cc}
 & 1\\
1
\end{array}\right),\ \sigma_{2}=\left(\begin{array}{cc}
 & -i\\
i
\end{array}\right),\ \sigma_{3}=\left(\begin{array}{cc}
1\\
 & -1
\end{array}\right).
\end{align*}

We fix our discussion in the region $\Delta_{r}>0$. The coframe is
\begin{align*}
    e^{0}&=\sqrt{\frac{\Delta_{r}}{U}}\left(dt-\frac{a\sin^{2}\theta}{\Xi}d\phi\right)\ ,\ e^{1}=\sqrt{\frac{U}{\Delta_{\theta}}}d\theta\\
    e^{2}&=\sqrt{\frac{\Delta_{\theta}}{U}}\sin\theta\left(adt-\frac{r^{2}+a^{2}}{\Xi}d\phi\right),\ e^{3}=\sqrt{\frac{U}{\Delta_{r}}}dr.
\end{align*}
with dual frame
\begin{align*}
    e_{0}&=\frac{r^{2}+a^{2}}{\sqrt{U\Delta_{r}}}\left(\partial_{t}+\frac{a\Xi}{r^{2}+a^{2}}\partial_{\phi}\right)\ ,\ e_{1}=\sqrt{\frac{\Delta_{r}}{U}}\partial_{r}, \\
    e_{2}&=\sqrt{\frac{\Delta_{\theta}}{U}}\partial_{\theta}\ ,\ e_{3}=-\frac{1}{\sqrt{U\Delta_{\theta}}}\left(a\sin\theta\partial_{t}+\frac{\Xi}{\sin\theta}\partial_{\phi}\right).
\end{align*}
With respect to the above coframe, it gives that
\begin{align*}
    de^{0}&=C_{10}^{0}e^{1}\wedge e^{0}+C_{30}^{0}e^{3}\wedge e^{0}+C_{12}^{0}e^{1}\wedge e^{2},\\
    de^{1}&=C_{31}^{1}e^{3}\wedge e^{1},\\
    de^{2}&=C_{30}^{2}e^{3}\wedge e^{0}+C_{32}^{2}e^{3}\wedge e^{2}+C_{12}^{2}e^{1}\wedge e^{2},\\
    de^{3}&=C_{31}^{3}e^{3}\wedge e^{1},
\end{align*}
where
\begin{align*}
	C_{10}^{0}&=-\frac{a^2}{U}\sqrt{\frac{\Delta_{\theta}}{U}}\sin\theta\cos\theta,\quad
    C_{30}^{0}=\partial_{r}\sqrt{\frac{\Delta_{r}}{U}},\quad
    C_{12}^{0}=\frac{2a}{U}\sqrt{\frac{\Delta_{r}}{U}}\cos\theta,\\
    C_{31}^{1}&=\frac{r}{U}\sqrt{\frac{\Delta_{r}}{U}},\quad
    C_{12}^{2}=\frac{1}{\sin\theta}\partial_{\theta}\left(\sqrt{\frac{\Delta_{\theta}}{U}}\sin\theta\right),\quad
    C_{32}^{2}=\frac{r}{U}\sqrt{\frac{\Delta_{r}}{U}},\quad\\
    C_{30}^{2}&=-\frac{2ar}{U}\sqrt{\frac{\Delta_{\theta}}{U}}\sin\theta,\quad
    C_{31}^{3}=\frac{a^2}{U}\sqrt{\frac{\Delta_{\theta}}{U}}\sin\theta\cos\theta.
\end{align*}
Thus connection 1-forms are
\begin{equation}\label{1-form}
\begin{aligned}
    \omega^{0}_{\ 1}&=-\omega_{01}=C_{10}^{0}e^{0}+\frac{1}{2}C_{12}^{0}e^{2},\quad
    \omega^{0}_{\ 2}=-\omega_{02}=-\frac{1}{2}C_{30}^{2}e^{3}-\frac{1}{2}C_{12}^{0}e^{1},\\
    \omega^{0}_{\ 3}&=-\omega_{03}=C_{30}^{0}e^{0}-\frac{1}{2}C_{30}^{2}e^{2},\quad
    \omega^{1}_{\ 2}=\omega_{12}=\frac{1}{2}C_{12}^{0}e^{0}-C_{12}^{2}e^{2},\\
    \omega^{1}_{\ 3}&=\omega_{13}=C_{31}^{3}e^{3}+C_{31}^{1}e^{1},\quad 
    \omega^{2}_{\ 3}=\omega_{23}=\frac{1}{2}C_{30}^{2}e^{0}+C_{32}^{2}e^{2}.
\end{aligned}
\end{equation}

The spin connection is defined as
\begin{align*}
\tilde{\nabla}_{X}\Psi=X(\Psi)-\frac{1}{4}\omega_{\mu \nu}(X)e^{\mu}\cdot e^{\nu}\cdot\Psi, 
\end{align*}
where $X$ is a vector, $\Phi$ is a spinor, $e^\mu \cdot$ is the Clifford multiplication. Therefore, using \eqref{1-form}, we obtain
\begin{align*}
    \tilde{\nabla}_{e_{0}}\Psi=&e_{0}\Psi-\frac{1}{2}\omega_{01}(e_{0})e^{0}\cdot e^{1}\cdot\Psi-\frac{1}{2}\omega_{03}(e_{0})e^{0}\cdot e^{3}\cdot\Psi\\
    &-\frac{1}{2}\omega_{12}(e_{0})e^{1}\cdot e^{2}\cdot\Psi-\frac{1}{2}\omega_{23}(e_{0})e^{2}\cdot e^{3}\cdot\Psi,\nonumber\\
    \tilde{\nabla}_{e_{1}}\Psi=&e_{1}\Psi-\frac{1}{2}\omega_{02}(e_{1})e^{0}\cdot e^{2}\cdot\Psi-\frac{1}{2}\omega_{12}(e_{1})e^{1}\cdot e^{2}\cdot\Psi,\\
    \tilde{\nabla}_{e_{2}}\Psi=&e_{2}\Psi-\frac{1}{2}\omega_{01}(e_{2})e^{0}\cdot e^{1}\cdot\Psi-\frac{1}{2}\omega_{03}(e_{2})e^{0}\cdot e^{3}\cdot\Psi\\
    &-\frac{1}{2}\omega_{12}(e_{2})e^{1}\cdot e^{2}\cdot\Psi-\frac{1}{2}\omega_{23}(e_{2})e^{2}\cdot e^{3}\cdot\Psi,\\
    \tilde{\nabla}_{e_{3}}\Psi=&e_{3}\Psi-\frac{1}{2}\omega_{02}(e_{3})e^{0}\cdot e^{2}\cdot\Psi-\frac{1}{2}\omega_{13}(e_{3})e^{1}\cdot e^{3}\cdot\Psi.
\end{align*}
Using Clifford multiplication \eqref{clifford}, they can be written as the matrix forms
\begin{align}
    \tilde{\nabla}_{e_{\mu}}\Psi=e_{\mu}\Psi+E_{\mu}\cdot\Psi, \quad E_{\mu}=-\frac{1}{2}\left(\begin{array}{cc}
\epsilon_{\mu} & 0\\
0 & -\epsilon_{\mu}^{h}
\end{array}\right),\label{spin-conn}
\end{align}
where $\epsilon_{\mu}^{h}$ is the Hermitian conjugate of $\epsilon_{\mu}$ and
\begin{equation*}
	\begin{aligned}
	\epsilon_{0} &= \left(C_{10}^{0}-\frac{i}{2}C_{30}^{2}\right)\sigma_{1}+\left(C_{30}^{0}-\frac{i}{2}C_{12}^{0}\right)\sigma_{3},\\
	\epsilon_{1} &= \left(-\frac{1}{2}C_{12}^{0}+iC_{31}^{1}\right)\sigma_{2},\\
	\epsilon_{2} &= \left(\frac{1}{2}C_{12}^{0}-iC_{32}^{2}\right)\sigma_{1}-\left(\frac{1}{2}C_{30}^{2}-iC_{12}^{2}\right)\sigma_{3},\\
	\epsilon_{3} &= \left(-\frac{1}{2}C_{30}^{2}+iC_{31}^{3}\right)\sigma_{2}.
\end{aligned}
\end{equation*}

\section{Time-periodic Majorana fermions}
\ls
In this section, we prove the nonexistence of time-periodic Majorana fermions in Kerr-Newman type spacetime when the electric or magnetic charge is nonzero.

Firstly we simplify the Dirac equation \eqref{eq:Diracequ} on metric \eqref{kerr} when $\Psi$ is given by \eqref{majorana}. The Dirac operator with electromagnetic potential $A$ is
\begin{align}
D=e^{\mu}\cdot\left(\tilde{\nabla}_{e_\mu}+iA(e_{\mu})\right). \label{Dirac eq KN}
\end{align}
Denote $\mathcal{J}=\text{diag}(I_{2\times2},\,-I_{2\times2})$. In terms of \eqref{A} and \eqref{spin-conn}, we obtain
\begin{align*}
e^{\mu} \cdot e_{\mu}(\Psi)
=&\sqrt{\frac{\Delta_r}{U}}e^3 \cdot S^{-1}E\partial_r\psi+\sqrt{\frac{\Delta_\theta}{U}}e^1\cdot S^{-1}E\partial_\theta\psi \\
&-i\frac{r^2+a^2}{\sqrt{U\Delta_{r}}} \left(\omega + \frac{a\Xi}{r^2+a^2}(k+\frac{1}{2})\right)e^0 \cdot \mathcal{J}S^{-1}E\psi \\
 &+\frac{1}{2U^{\frac{3}{2}}}e^3 \cdot \Big(S-\partial_r(U\sqrt{\Delta_r})S^{-1}\Big) E\psi\\
 &+\frac{a\sin\theta}{2U^\frac{3}{2}}\sqrt{\frac{\Delta_\theta}{\Delta_r}}e^1 \cdot \left(i\mathcal{J}S+2a\cos\theta\sqrt{\Delta_r}S^{-1}\right)E\psi\\
 &+\frac{i}{\sqrt{U\Delta_\theta}}\left(a\omega\sin\theta+\frac{\Xi}{\sin\theta}(k+\frac{1}{2})\right)e^2 \cdot S^{-1}\mathcal{J}E\psi,\\
e^\mu\cdot E_\mu\Psi =&\frac{2\Delta_\theta-\Xi}{2\sqrt{U\Delta_\theta}} \cot\theta e^1 \cdot S^{-1} E\psi-\frac{ia}{2} \sqrt{\frac{\Delta_r\Delta_\theta}{U}}\sin\theta e^1 \cdot \mathcal{J}S^{-3}E\psi\\
 &+\frac{\partial_r \sqrt{\Delta_r}}{2\sqrt{U}}e^3 \cdot S^{-1}E\psi+\frac{\Delta_r}{2\sqrt{U}} e^3 \cdot S^{-3}E\psi,\\
e^\mu \cdot \Big(iA(e_\mu)\Big)\Psi=&-\frac{iQr}{\sqrt{U\Delta_r}} e^0 \cdot S^{-1}E\psi-\frac{iP\cot\theta}{\sqrt{U\Delta_\theta}} e^2 \cdot S^{-1}E\psi.
\end{align*}
Note that
\begin{align*}
	&e^\mu\mathcal{J}=-\mathcal{J}e^\mu,\quad
	e^\mu E=E^{-1}e^\mu, \quad
	e^\mu E^{-1}=Ee^\mu,\\
	&e^\mu S^{-1}=\frac{1}{\sqrt{U\Delta_r}}Se^\mu,\quad
	e^\mu S=\sqrt{U\Delta_r}S^{-1}e^\mu.
\end{align*}
Substituting the above formulas into \eqref{Dirac eq KN}, we obtain 
\begin{equation}
D\Psi=\frac{1}{U\sqrt{\Delta_r}}SE^{-1}\left(\sqrt{\Delta_r}\mathcal{D}_r-\sqrt{\Delta_\theta}\mathcal{L}_\theta \right)\psi \label{Dirac eq KN-1}
\end{equation}
with
\begin{align*}
\mathcal{D}_r=&e^{3}\partial_r+\frac{i}{\Delta_r}\bigg(\omega(r^2+a^2)+a\Xi(k+\frac{1}{2})\bigg)\mathcal{J}e^0-\frac{iQr}{\Delta_r}e^0,\\
\mathcal{L}_\theta=&-e^{1}\partial_\theta+\frac{i}{\Delta_\theta}\bigg(\omega a\sin\theta+\frac{\Xi}{\sin\theta}(k+\frac{1}{2})\bigg)\mathcal{J}e^2\\
	               &-\Big(1-\frac{\Xi}{2\Delta_\theta}\Big)\cot\theta e^1+\frac{iP}{\Delta_\theta}\cot\theta e^2.
\end{align*}
Denote $\lambda_{\omega k}=\lambda e^{-2i(\omega t +(k+\frac{1}{2})\phi)}$. Using \eqref{Dirac eq KN-1}, we can reduce the Dirac equation \eqref{eq:Diracequ} to
\begin{equation}\label{Maj1}
	D_r \psi = L_\theta \psi,\quad  \psi=\left(\begin{array}{c}
R_{-}(r)\Theta_{-}(\theta)\\
R_{+}(r)\Theta_{+}(\theta)\\
\bar{R}_{+}(r)\bar{\Theta}_{+}(\theta)\\
-\bar{R}_{-}(r)\bar{\Theta}_{-}(\theta)
\end{array}\right)
\end{equation}
where
\begin{align*}
	D_r &= \left(\begin{array}{cccc}
		-i\lambda_{\omega k} r &  & \sqrt{\Delta_{r}}D_{r.00}\\
		& -i\lambda_{\omega k} r &  & \sqrt{\Delta_{r}}D_{r,01}\\
		\sqrt{\Delta_{r}}D_{r,11} &  & -i \overline{\lambda_{\omega k}} r\\
 		& \sqrt{\Delta_{r}}D_{r,10} &  & -i \overline{\lambda_{\omega k}} r
	\end{array}\right),\\
	L_\theta &=\left(\begin{array}{cccc}
		a\lambda_{\omega k}\cos\theta &  &  & \sqrt{\Delta_{\theta}}L_{\theta,00}\\
 		& a\lambda_{\omega k}\cos\theta & \sqrt{\Delta_{\theta}}L_{\theta,01}\\
 		& \sqrt{\Delta_{\theta}}L_{\theta,11} & -a\overline{\lambda_{\omega k}}\cos\theta\\
		\sqrt{\Delta_{\theta}}L_{\theta,10} &  &  & -a\overline{\lambda_{\omega k}}\cos\theta
	\end{array}\right)
\end{align*}
and, for $l$, $m$=0, 1,
\begin{align*}
	D_{r,lm} =& (-1)^m\partial_r+(-1)^{l}\frac{i}{\Delta_r}\left(\omega(r^2+a^2)+(k+\frac{1}{2})\Xi a\right)-\frac{iQr}{\Delta_r},\\
	L_{\theta,lm} =& -(-1)^l \partial_\theta + \frac{(-1)^{l+m}}{\Delta_\theta}\bigg(\omega a \sin\theta+\frac{\Xi}{\sin\theta}(k+\frac{1}{2})\\
                   &+(-1)^l P\cot\theta-(-1)^m \Big(\Delta_\theta-\frac{\Xi}{2} \Big)\cot\theta \bigg).
\end{align*}
Writing down each row of \eqref{Maj1}, we get
\begin{align}
-i\lambda_{\omega k} rR_{-}\Theta_{-}+\sqrt{\Delta_{r}}D_{r,00}\bar{R}_{+}\bar{\Theta}_{+}
        &=a\lambda_{\omega k}\cos\theta R_{-}\Theta_{-}-\sqrt{\Delta_\theta}L_{\theta,00}\bar{R}_{-}\bar{\Theta}_{-},
        \label{eq:majequ1}\\
-i\lambda_{\omega k} rR_{+}\Theta_{+}-\sqrt{\Delta_{r}}D_{r,01}\bar{R}_{-}\bar{\Theta}_{-}
        &= a\lambda_{\omega k}\cos\theta R_{+}\Theta_{+}+\sqrt{\Delta_{\theta}}L_{\theta,01}\bar{R}_{+}\bar{\Theta}_{+},
        \label{eq:majequ2}\\
-i\overline{\lambda_{\omega k}} r\bar{R}_{+}\bar{\Theta}_{+}+\sqrt{\Delta_{r}}D_{r,11}R_{-}\Theta_{-}
        &=-a\overline{\lambda_{\omega k}}\cos\theta\bar{R}_{+}\bar{\Theta}_{+}+\sqrt{\Delta_{\theta}}L_{\theta,11}R_{+}\Theta_{+},
        \label{eq:majequ3}\\
i\overline{\lambda_{\omega k}} r\bar{R}_{-}\bar{\Theta}_{-}+\sqrt{\Delta_{r}}D_{r,10}R_{+}\Theta_{+}
        &=a\overline{\lambda_{\omega k}}\cos\theta\bar{R}_{-}\bar{\Theta}_{-}+\sqrt{\Delta_{\theta}}L_{\theta,10}R_{-}\Theta_{-}.
        \label{eq:majequ4}
\end{align}
These equations can not separate into radial and angular equations. But
\begin{align*}
    \overline{D}_{r,lm}=-D_{r,l\bar{m}},\quad \overline{L}_{\theta,lm}=L_{\theta,lm}
\end{align*}
and
\begin{align*}
	D_{r,lm}+D_{r,\bar{l}\bar{m}} = -\frac{2iQr}{\Delta_r},\quad
	L_{\theta,\bar{l}m}+L_{\theta,lm} = 2(-1)^{m}\frac{P\cot\theta}{\Delta_\theta},
\end{align*}
where $\bar{l}=(l+1)\mod 2$ and $\bar{m}=(m+1)\mod 2$. Thus, subtracting the complex conjugation of \eqref{eq:majequ1} from \eqref{eq:majequ4}, adding the complex conjugation of \eqref{eq:majequ2} with \eqref{eq:majequ3} respectively, we obtain two algebraic identities
\begin{align*}
    i\alpha(r)R_{-}\Theta_{-}-\beta(\theta)R_{+}\Theta_{+}=0,\quad 
	\beta(\theta)R_{-}\Theta_{-}+i\alpha(r)R_{+}\Theta_{+}=0,
\end{align*}
where
\begin{align*}
	\alpha(r)=\frac{Qr}{\sqrt{\Delta_r}},\quad \beta(\theta)=\frac{P\cot\theta}{\sqrt{\Delta_\theta}}.
\end{align*}
Therefore
\begin{align*}
\Big(\alpha(r) ^2 -\beta (\theta) ^2 \Big)R_{+}\Theta_{+}=\Big(\alpha(r) ^2 -\beta (\theta) ^2 \Big)R_{-}\Theta_{-}=0.
\end{align*}
If $R_{+}\Theta_{+}$ or $R_{-}\Theta_{-}$ is nontrival, then it must hold that $\alpha(r)=\pm \beta(\theta)$. As $\alpha(r)$ depends only on $r>0$ (outside event horizon), and $\beta(\theta)$ depends only on $\theta$, then both are constant. Therefore three cases occur: (i) $P=Q=0$; (ii) $P\neq 0$, $Q=0$, $\theta =\frac{\pi}{2}$; (iii) $P \neq 0$, $Q\neq 0$, $r=r_0$ is a positive constant, $\theta=\theta_0$ is a constant. But $R_{+}\Theta_{+}=R_{-}\Theta_{-}=0$ outside hypersurface $\theta =\frac{\pi}{2}$ equipped with the metric
\begin{align*}
    ds_{3}^{2}=&-\frac{\Delta_{r}}{r^2}\left(dt-\frac{a}{\Xi}d\phi\right)^{2}
    +\frac{r^2}{\Delta_{r}}dr^{2}
    +\frac{1}{r^2}\left(adt-\frac{r^{2}+a^{2}}{\Xi}d\phi\right)^{2}
\end{align*}
in case (ii), and outside 2-surface equipped with the metric
\begin{align*}
    ds_{2} ^{2}=-\frac{\Delta_{r}(r_0)}{U(r_0, \theta_0)}\left(dt-\frac{a\sin^{2}\theta_0}{\Xi}d\phi\right)^{2}
                +\frac{\Delta_{\theta}(\theta_0)\sin^{2}\theta_0}{U(r_0, \theta_0)}\left(a dt-\frac{r_0 ^{2}+a^{2}}{\Xi}d\phi\right)^{2}
\end{align*}
in case (iii). This indicates that Majarana fermions are not differentiable in cases (ii) and (iii). Therefore, if $P\neq 0$ or $Q\neq 0$, we conclude that there is no differentiable time-periodic Majorana fermions in Kerr-Newman-type spacetimes.

\section{Conclusion}
\ls

We point out that Chandrasekhar's separation for time-period Dirac fermions is not consistent with the condition for Majorana fermions, and we introduce new separation for time-period Majorana fermions. With this separation, the Dirac equation can not be transformed into radial and angular equations, which is done for Chandrasekhar's separation. But it is separated into four differential equations, which yield two algebraic identities. When the electric or magnetic charge is nonzero, they conclude that there is no differentiable time-periodic Majorana fermions outside the event horizon in Kerr-Newman and Kerr-Newman-AdS spacetimes, or between the event horizon and the cosmological horizon in Kerr-Newman-dS spacetime. This conclusion plays a role for searching free Majorana fermions when gravitational effect is considered. 

\bigskip

\footnotesize {

\noindent {\bf Acknowledgement} The authors are grateful to the referees for many valuable suggestions. The work is supported by the National Natural Science Foundation of China 12326602.

}

\end{document}